\documentclass[manuscript]{acmart}

\AtBeginDocument{%
  \providecommand\BibTeX{{%
    \normalfont B\kern-0.5em{\scshape i\kern-0.25em b}\kern-0.8em\TeX}}}

\setcopyright{acmcopyright}
\copyrightyear{2023}
\acmYear{2023}
\acmDOI{XXXXXXX.XXXXXXX}

\begin{document}

\title{Exploring Gender Biases in Language Patterns of Human-Conversational Agent Conversations}

\author{Weizi Liu}
\email{weizil2@illinois.edu}
\affiliation{%
  \institution{University of Illinois at Urbana-Champaign}
  \streetaddress{614 E. Daniel St., Room 4116}
  \city{Champaign}
  \state{IL}
  \country{USA}
  \postcode{61820}
}

\begin{abstract}
  With the rise of human-machine communication, machines are increasingly designed with humanlike characteristics, such as gender, which can inadvertently trigger cognitive biases. Many conversational agents (CAs), such as voice assistants and chatbots, are predominantly designed with female personas. This leads to concerns about perpetuating gender stereotypes and inequality. Critiques have emerged regarding the potential objectification of females and the reinforcement of gender stereotypes by these technologies. This research, situated in conversational AI design, aims to delve deeper into the impacts of gender biases in human-CA interactions. From a behavioral and communication research standpoint, this program focuses not only on perceptions but also the linguistic styles of users when interacting with CAs, as previous research has rarely explored. It aims to understand how pre-existing gender biases might be triggered by CAs' gender designs. It further investigates how CAs' gender designs may reinforce gender biases and extend them to human-human communication. The findings aim to inform the ethical design of conversational agents, addressing whether gender assignment in CAs is appropriate and how to promote gender equality in design.
\end{abstract}



\maketitle

\section{Introduction}
Communication between humans and machines is becoming ubiquitous. Designers are making machine agents more sophisticated with humanlike characteristics and identities (e.g., race, gender, nationality, etc.). Consequently, this can trigger cognitive biases linked to these attributes. Gender biases stand out as a prominent example. Conversational agents (CAs), such as voice assistants and text-based chatbots, are predominantly depicted as friendly, obedient, and submissive females\cite{costa_conversing_2018}. For example, Siri, Alexa, and Cortana were primarily or by default female-voiced \cite{chin_how_2020}; Samsung’s virtual assistant is embodied in a female avatar. The gendered technological design has sparked public and scholarly criticism. UNESCO warned about sexism in technology\cite{west_id_2019}; critical scholars also argued that “gendered” technologies could exacerbate female objectification, gender stereotypes, and gender power imbalance in the era of artificial intelligence \cite{bergen_id_2016}\cite{poster_5_2016}. Besides critiques, more empirical research is needed to examine how such technology design penetrates our daily lives, cultivates norms, and gradually shapes culture. A potential concerning consequence could be that if users come to expect female CAs to be submissive—despite their intended role as equal collaborators—this perception could become normalized in daily interactions. Over time, this might inadvertently encourage users to adopt a more dominant stance towards women in real-world scenarios. Situated in conversational AI design, I propose a working research program \cite{liu2023gender} to explore the gender biases in human-CA communication impacted by CAs that are designed with gendered humanlike characteristics. With a particular focus on language use in conversations, this research initiative seeks to understand how users project prevailing gender biases onto gendered CAs in their ways of talking to them and how such biases might be reinforced through the design of gendered CAs. The findings of this research are expected to answer practical questions in conversational design: is a gendered CA appropriate? When would it be beneficial or detrimental to assign a gender to a CA? Drawing on the observed cognitive and behavioral patterns, how should we mitigate biases and promote gender equality in CA design?

\section{Related Work}
Gender biases in technology have raised increasing attention in recent years. Existing research has focused on tackling gender biases in data and algorithms\cite{bozdag_bias_2013}, system design\cite{vorvoreanu_gender_2019}, and the reference and representation of participants\cite{bradley_gendered_2015}\cite{offenwanger_diagnosing_2021}, etc. Not enough work has explored user behavior impacted by gender biases when interacting with technology extensively. Based on theories in social identity, interpersonal communication, and human-computer interaction, this research program explores gender biases in user evaluation and communicative behavior in conversations with a gendered CA. Gender and sex roles are important social identities that influence language use, social perception, and stereotyping in human communication processes \cite{leaper_meta-analytic_2007}\cite{palomares_explaining_2008}. For example, feminine language tends to be affiliative and hedging; masculine language tends to be assertive and commanding. People adjust their communicative behavior to accommodate or differentiate themselves from their communication partners (see communication accommodation theory; \cite{gallois2005communication}), which is salient in gender-based communication \cite{boggs1999canary}\cite{namy_gender_2002}. In human-computer interaction, researchers explored how humans automatically apply interpersonal norms to interactions with machines (see “Computers Are Social Actors” paradigm; \cite{reeves_media_1996}), which include gender stereotypes that were reflected by credibility and attraction evaluations of machines \cite{lee2000can}\cite{nass_machines_2000}. Furthermore, recent scholarship pointed out that human-machine communication may possess unique characteristics, norms, and social scripts \cite{gambino_building_2020}\cite{lombard_social_2021}, which inspires the comparison between the gendered human-machine communication schema and the interpersonal communication schema. 

Early studies based on CASA initiated the exploration of gender effects in human-AI communication. For example, \cite{lee2000can} found that a computer with a female voice was associated with greater attractiveness yet less opinion influence than one with a male voice. In more recent research, researchers found the majority of existing chatbot designs are more female than male\cite{feine_gender_2019}. However, users do not seem to show a clear consensus on gender preference and the authors therefore recommended customizing chatbots' gender for individual users\cite{curry2020conversational}. Besides design concepts and user attitudes, research on user behavior in human-CA communication has focused on deviant behaviors such as aggression and abuse toward a female CA \cite{de2006sex}. However, gender effects in ordinary and non-extreme user-CA conversations may have indirect and implicit influence \cite{adams_addressing_2019}, thus revealing more research opportunities. While the CASA paradigm offers a general framework, more specific processes need to be identified for a fuller understanding of the issue. Moreover, although existing research on human-CA communication has referred to many interpersonal theories and perspectives, few studies applied interpersonal communication research methods to study human-CA communication. This research program incorporates dyadic observational methods to observe and compare gender biases from the sentiment, dynamics, and interactional patterns in human-CA conversations.

\section{Research Plan}

I aim to answer three sets of overarching questions: a) How do gender identities influence human-CA communication? Put differently, do the mere gender identities of a CA alter users' attitudes and the manner in which users communicate with the CA? b) What are the differences between the gendered human-CA communication schema and the gendered interpersonal communication schema? Are human-CA conversations more gender-neutral or gender-biased? c) How will gendered human-CA communication influence subsequent human-human communication processes, more specifically, mitigate or reinforce gender stereotypical perceptions and gendered language use? 
I propose a mixed-methods approach to explore the role of gender in human-CA conversations to understand the issue comprehensively. First, I propose an observational study of unrestricted interactions between human users and a female- or male-voiced CA in a laboratory and qualitatively analyze the conversations to detect gendered communication cues and patterns, including: a) gender-linked language use (e.g., affiliation, assertiveness, hedges, etc.), b) interactional patterns (e.g., engagement, interruptions, convergence, divergence, etc.), and c) sentiment (valence, emotions, and tone of voice). 

Second, I propose an online experiment aimed at quantifying the patterns identified in the first step. This experiment will examine how various dyads—comprising a user and a gendered (female, male, or neutral) CA—differ in their expectations and evaluations of the CA, including factors like credibility and attractiveness. Additionally, the experiment will assess the language users employ when interacting with the CA. I propose a text analysis on gendered human-CA chat history using dictionaries such as LIWC (and more advanced natural language processing tools), with a focus on emotions, tones, femininity, masculinity, power dynamics, etc. The preliminary findings included many details such as that users carried over human communication norms and habits to human-machine communication, with females tending to be more cooperative and detail-oriented and males tending to be less emotional and more concise. A female CA was preferred as an option and more likely to be anthropomorphized, but was evaluated less favorably than a male CA, which could possibly be attributed to novelty effect of a male CA and the violation of heightened expectation in a female CA. Dyads consisting of male users and male CAs exhibited strong rapport, as evidenced by high evaluations, engagement, and frequent use of power-related language\cite{liu2023gender}.

The final step is to examine if communicating with a CA may influence subsequent human-human interactions, in other words, if users carry over their communication schema with machines to humans, thus reducing or enhancing gender biases. One assumption is that if users command a female agent as an assistant and internalize this role, they may talk less respectfully to females in real life. I plan to design a lab experiment to test users’ attitudes and behavioral change after interacting with a gender-stereotypical CA (e.g., a female agent as a personal assistant) or a gender counter-stereotypical CA (e.g., a female agent as a tech expert) in a certain collaborative role. The experiment will be 3 (CA’s gender: female vs. male vs. neutral) * 2 (CA’s role: a personal assistant vs. a tech expert). Participants will be asked to: a) have a conversation with one of the CAs to complete a task collaboratively, b) evaluate their experience, and c) chat with a female human agent. I will test how evaluations of the CA, changes in gender-related attitudes, and communication patterns with the subsequent human agent differ across conditions. 

\section{Discussion}
Given the expected gender biases discovered in human-CA conversations, it is important to consider whether the advantages of creating a humanlike persona for a conversational agent (CA) are worth the potential risk of perpetuating biases. One might question whether assigning a gender, especially a female one, should be prohibited due to concerns about sexualization and biases. This raises the dilemma: should we abandon utilizing positive attributes often associated with female identities, like rapport-building and empathy\cite{stumpf2020gender}? When incorporating gender identities into design, it is crucial to be considerate of the context and the relationship dynamics between users and CAs. The innovation of this research lies in proposing an agenda for theory building of HCI research regarding gender biases and practicing a dyadic approach from interpersonal communication to study human-AI communication. It elicits extensive discussion on an important social issue related to gender equality in technology and provides implications for mindful and ethical conversational AI design.

\bibliographystyle{ACM-Reference-Format}
\bibliography{ref1}


\begin{thebibliography}{25}


\ifx \showCODEN    \undefined \def \showCODEN     #1{\unskip}     \fi
\ifx \showDOI      \undefined \def \showDOI       #1{#1}\fi
\ifx \showISBNx    \undefined \def \showISBNx     #1{\unskip}     \fi
\ifx \showISBNxiii \undefined \def \showISBNxiii  #1{\unskip}     \fi
\ifx \showISSN     \undefined \def \showISSN      #1{\unskip}     \fi
\ifx \showLCCN     \undefined \def \showLCCN      #1{\unskip}     \fi
\ifx \shownote     \undefined \def \shownote      #1{#1}          \fi
\ifx \showarticletitle \undefined \def \showarticletitle #1{#1}   \fi
\ifx \showURL      \undefined \def \showURL       {\relax}        \fi
\providecommand\bibfield[2]{#2}
\providecommand\bibinfo[2]{#2}
\providecommand\natexlab[1]{#1}
\providecommand\showeprint[2][]{arXiv:#2}

\bibitem[Adams and Ni~Loideain(2019)]%
        {adams_addressing_2019}
\bibfield{author}{\bibinfo{person}{Rachel Adams} {and} \bibinfo{person}{Nora
  Ni~Loideain}.} \bibinfo{year}{2019}\natexlab{}.
\newblock \bibinfo{title}{Addressing {Indirect} {Discrimination} and {Gender}
  {Stereotypes} in {AI} {Virtual} {Personal} {Assistants}: {The} {Role} of
  {International} {Human} {Rights} {Law}}.
\newblock
\newblock
\urldef\tempurl%
\url{https://doi.org/10.2139/ssrn.3392243}
\showDOI{\tempurl}


\bibitem[Bergen(2016)]%
        {bergen_id_2016}
\bibfield{author}{\bibinfo{person}{Hilary Bergen}.}
  \bibinfo{year}{2016}\natexlab{}.
\newblock \showarticletitle{‘{I}’d {Blush} if {I} {Could}’: {Digital}
  {Assistants}, {Disembodied} {Cyborgs} and the {Problem} of {Gender}}.
\newblock
\urldef\tempurl%
\url{https://www.semanticscholar.org/paper/%E2%80%98I%E2%80%99d-Blush-if-I-Could%E2%80%99%3A-Digital-Assistants%2C-Cyborgs-Bergen/1f00770d4ef6a27c4c1abedc11915bdbf4c64b4b}
\showURL{%
\tempurl}


\bibitem[Boggs and giles(1999)]%
        {boggs1999canary}
\bibfield{author}{\bibinfo{person}{Cathy Boggs} {and} \bibinfo{person}{Howard
  giles}.} \bibinfo{year}{1999}\natexlab{}.
\newblock \showarticletitle{‘The canary in the coalmine’: The
  nonaccommodation cycle in the gendered workplace}.
\newblock \bibinfo{journal}{\emph{International journal of applied
  linguistics}} \bibinfo{volume}{9}, \bibinfo{number}{2}
  (\bibinfo{year}{1999}), \bibinfo{pages}{223--245}.
\newblock


\bibitem[Bozdag(2013)]%
        {bozdag_bias_2013}
\bibfield{author}{\bibinfo{person}{Engin Bozdag}.}
  \bibinfo{year}{2013}\natexlab{}.
\newblock \showarticletitle{Bias in algorithmic filtering and personalization}.
\newblock \bibinfo{journal}{\emph{Ethics and Information Technology}}
  \bibinfo{volume}{15}, \bibinfo{number}{3} (\bibinfo{date}{Sept.}
  \bibinfo{year}{2013}), \bibinfo{pages}{209--227}.
\newblock
\showISSN{1572-8439}
\urldef\tempurl%
\url{https://doi.org/10.1007/s10676-013-9321-6}
\showDOI{\tempurl}


\bibitem[Bradley et~al\mbox{.}(2015)]%
        {bradley_gendered_2015}
\bibfield{author}{\bibinfo{person}{Adam Bradley}, \bibinfo{person}{Cayley
  MacArthur}, \bibinfo{person}{Mark Hancock}, {and} \bibinfo{person}{Sheelagh
  Carpendale}.} \bibinfo{year}{2015}\natexlab{}.
\newblock \showarticletitle{Gendered or neutral? considering the language of
  {HCI}}. In \bibinfo{booktitle}{\emph{Proceedings of the 41st {Graphics}
  {Interface} {Conference}}} \emph{(\bibinfo{series}{{GI} '15})}.
  \bibinfo{publisher}{Canadian Information Processing Society},
  \bibinfo{address}{CAN}, \bibinfo{pages}{163--170}.
\newblock
\showISBNx{978-0-9947868-0-7}


\bibitem[Chin and Robison(2020)]%
        {chin_how_2020}
\bibfield{author}{\bibinfo{person}{Caitlin Chin} {and}
  \bibinfo{person}{Mishaela Robison}.} \bibinfo{year}{2020}\natexlab{}.
\newblock \showarticletitle{How {AI} bots and voice assistants reinforce gender
  bias}.
\newblock  (\bibinfo{date}{Nov.} \bibinfo{year}{2020}).
\newblock
\urldef\tempurl%
\url{https://policycommons.net/artifacts/4143566/how-ai-bots-and-voice-assistants-reinforce-gender-bias/4952630/}
\showURL{%
\tempurl}
\newblock
\shownote{Publisher: Brookings Institution}.


\bibitem[Costa(2018)]%
        {costa_conversing_2018}
\bibfield{author}{\bibinfo{person}{Pedro Costa}.}
  \bibinfo{year}{2018}\natexlab{}.
\newblock \showarticletitle{Conversing with personal digital assistants: on
  gender and artificial intelligence}.
\newblock \bibinfo{journal}{\emph{Journal of Science and Technology of the
  Arts}} (\bibinfo{date}{Sept.} \bibinfo{year}{2018}), \bibinfo{pages}{59--72
  Páginas}.
\newblock
\urldef\tempurl%
\url{https://doi.org/10.7559/CITARJ.V10I3.563}
\showDOI{\tempurl}
\newblock
\shownote{Artwork Size: 59-72 Páginas Publisher: Journal of Science and
  Technology of the Arts}.


\bibitem[Curry et~al\mbox{.}(2020)]%
        {curry2020conversational}
\bibfield{author}{\bibinfo{person}{Amanda~Cercas Curry}, \bibinfo{person}{Judy
  Robertson}, {and} \bibinfo{person}{Verena Rieser}.}
  \bibinfo{year}{2020}\natexlab{}.
\newblock \showarticletitle{Conversational assistants and gender stereotypes:
  Public perceptions and desiderata for voice personas}. In
  \bibinfo{booktitle}{\emph{Proceedings of the second workshop on gender bias
  in natural language processing}}. \bibinfo{pages}{72--78}.
\newblock


\bibitem[De~Angeli et~al\mbox{.}(2006)]%
        {de2006sex}
\bibfield{author}{\bibinfo{person}{Antonella De~Angeli},
  \bibinfo{person}{Sheryl Brahnam}, {et~al\mbox{.}}}
  \bibinfo{year}{2006}\natexlab{}.
\newblock \showarticletitle{Sex stereotypes and conversational agents}.
\newblock \bibinfo{journal}{\emph{Proc. of Gender and Interaction: real and
  virtual women in a male world, Venice, Italy}} (\bibinfo{year}{2006}).
\newblock


\bibitem[Feine et~al\mbox{.}(2019)]%
        {feine_gender_2019}
\bibfield{author}{\bibinfo{person}{Jasper Feine}, \bibinfo{person}{Ulrich
  Gnewuch}, \bibinfo{person}{Stefan Morana}, {and} \bibinfo{person}{Alexander
  Maedche}.} \bibinfo{year}{2019}\natexlab{}.
\newblock \showarticletitle{Gender {Bias} in {Chatbot} {Design}}. In
  \bibinfo{booktitle}{\emph{Chatbot {Research} and {Design}: {Third}
  {International} {Workshop}, {CONVERSATIONS} 2019, {Amsterdam}, {The}
  {Netherlands}, {November} 19–20, 2019, {Revised} {Selected} {Papers}}}.
  \bibinfo{publisher}{Springer-Verlag}, \bibinfo{address}{Berlin, Heidelberg},
  \bibinfo{pages}{79--93}.
\newblock
\showISBNx{978-3-030-39539-1}
\urldef\tempurl%
\url{https://doi.org/10.1007/978-3-030-39540-7_6}
\showDOI{\tempurl}


\bibitem[Gallois et~al\mbox{.}(2005)]%
        {gallois2005communication}
\bibfield{author}{\bibinfo{person}{Cindy Gallois}, \bibinfo{person}{Tania
  Ogay}, {and} \bibinfo{person}{Howard Giles}.}
  \bibinfo{year}{2005}\natexlab{}.
\newblock \showarticletitle{Communication accommodation theory: A look back and
  a look ahead}.
\newblock In \bibinfo{booktitle}{\emph{Theorizing about intercultural
  communication}}. \bibinfo{publisher}{Thousand Oaks: Sage},
  \bibinfo{pages}{121--148}.
\newblock


\bibitem[Gambino et~al\mbox{.}(2020)]%
        {gambino_building_2020}
\bibfield{author}{\bibinfo{person}{Andrew Gambino}, \bibinfo{person}{Jesse
  Fox}, {and} \bibinfo{person}{Rabindra Ratan}.}
  \bibinfo{year}{2020}\natexlab{}.
\newblock \showarticletitle{Building a {Stronger} {CASA}: {Extending} the
  {Computers} {Are} {Social} {Actors} {Paradigm}}. In
  \bibinfo{booktitle}{\emph{Human-{Machine} {Communication}}},
  Vol.~\bibinfo{volume}{1}. \bibinfo{pages}{71--86}.
\newblock
\urldef\tempurl%
\url{https://doi.org/10.30658/hmc.1.5}
\showDOI{\tempurl}
\newblock
\shownote{ISSN: 2638-6038, 2638-602X Journal Abbreviation: HMC}.


\bibitem[Leaper and Ayres(2007)]%
        {leaper_meta-analytic_2007}
\bibfield{author}{\bibinfo{person}{Campbell Leaper} {and}
  \bibinfo{person}{Melanie~M. Ayres}.} \bibinfo{year}{2007}\natexlab{}.
\newblock \showarticletitle{A {Meta}-{Analytic} {Review} of {Gender}
  {Variations} in {Adults}' {Language} {Use}: {Talkativeness}, {Affiliative}
  {Speech}, and {Assertive} {Speech}}.
\newblock \bibinfo{journal}{\emph{Personality and Social Psychology Review}}
  \bibinfo{volume}{11}, \bibinfo{number}{4} (\bibinfo{date}{Nov.}
  \bibinfo{year}{2007}), \bibinfo{pages}{328--363}.
\newblock
\showISSN{1088-8683}
\urldef\tempurl%
\url{https://doi.org/10.1177/1088868307302221}
\showDOI{\tempurl}
\newblock
\shownote{Publisher: SAGE Publications Inc}.


\bibitem[Lee et~al\mbox{.}(2000)]%
        {lee2000can}
\bibfield{author}{\bibinfo{person}{Eun~Ju Lee}, \bibinfo{person}{Clifford
  Nass}, {and} \bibinfo{person}{Scott Brave}.} \bibinfo{year}{2000}\natexlab{}.
\newblock \showarticletitle{Can computer-generated speech have gender? An
  experimental test of gender stereotype}. In \bibinfo{booktitle}{\emph{CHI'00
  extended abstracts on Human factors in computing systems}}.
  \bibinfo{pages}{289--290}.
\newblock


\bibitem[Liu and Yao(2023)]%
        {liu2023gender}
\bibfield{author}{\bibinfo{person}{Weizi Liu} {and} \bibinfo{person}{Mike
  Yao}.} \bibinfo{year}{2023}\natexlab{}.
\newblock \showarticletitle{Gender identity and influence in human-machine
  communication: A mixed-methods exploration}.
\newblock \bibinfo{journal}{\emph{Computers in Human Behavior}}
  \bibinfo{volume}{144} (\bibinfo{year}{2023}), \bibinfo{pages}{107750}.
\newblock


\bibitem[Lombard and Xu(2021)]%
        {lombard_social_2021}
\bibfield{author}{\bibinfo{person}{Matthew Lombard} {and} \bibinfo{person}{Kun
  Xu}.} \bibinfo{year}{2021}\natexlab{}.
\newblock \showarticletitle{Social {Responses} to {Media} {Technologies} in the
  21st {Century}: {The} {Media} are {Social} {Actors} {Paradigm}}.
\newblock \bibinfo{journal}{\emph{Human-Machine Communication}}
  \bibinfo{volume}{2}, \bibinfo{number}{1} (\bibinfo{date}{April}
  \bibinfo{year}{2021}).
\newblock
\showISSN{2638-6038}
\urldef\tempurl%
\url{https://doi.org/10.30658/hmc.2.2}
\showDOI{\tempurl}


\bibitem[Namy et~al\mbox{.}(2002)]%
        {namy_gender_2002}
\bibfield{author}{\bibinfo{person}{Laura~L. Namy}, \bibinfo{person}{Lynne~C.
  Nygaard}, {and} \bibinfo{person}{Denise Sauerteig}.}
  \bibinfo{year}{2002}\natexlab{}.
\newblock \showarticletitle{Gender {Differences} in {Vocal} {Accommodation}::
  {The} {Role} of {Perception}}.
\newblock \bibinfo{journal}{\emph{Journal of Language and Social Psychology}}
  \bibinfo{volume}{21}, \bibinfo{number}{4} (\bibinfo{date}{Dec.}
  \bibinfo{year}{2002}), \bibinfo{pages}{422--432}.
\newblock
\showISSN{0261-927X}
\urldef\tempurl%
\url{https://doi.org/10.1177/026192702237958}
\showDOI{\tempurl}
\newblock
\shownote{Publisher: SAGE Publications Inc}.


\bibitem[Nass and Moon(2000)]%
        {nass_machines_2000}
\bibfield{author}{\bibinfo{person}{Clifford Nass} {and}
  \bibinfo{person}{Youngme Moon}.} \bibinfo{year}{2000}\natexlab{}.
\newblock \showarticletitle{Machines and {Mindlessness}: {Social} {Responses}
  to {Computers}}.
\newblock \bibinfo{journal}{\emph{Journal of Social Issues}}
  \bibinfo{volume}{56}, \bibinfo{number}{1} (\bibinfo{date}{Jan.}
  \bibinfo{year}{2000}), \bibinfo{pages}{81--103}.
\newblock
\showISSN{0022-4537, 1540-4560}
\urldef\tempurl%
\url{https://doi.org/10.1111/0022-4537.00153}
\showDOI{\tempurl}


\bibitem[Offenwanger et~al\mbox{.}(2021)]%
        {offenwanger_diagnosing_2021}
\bibfield{author}{\bibinfo{person}{Anna Offenwanger},
  \bibinfo{person}{Alan~John Milligan}, \bibinfo{person}{Minsuk Chang},
  \bibinfo{person}{Julia Bullard}, {and} \bibinfo{person}{Dongwook Yoon}.}
  \bibinfo{year}{2021}\natexlab{}.
\newblock \showarticletitle{Diagnosing {Bias} in the {Gender} {Representation}
  of {HCI} {Research} {Participants}: {How} it {Happens} and {Where} {We}
  {Are}}. In \bibinfo{booktitle}{\emph{Proceedings of the 2021 {CHI}
  {Conference} on {Human} {Factors} in {Computing} {Systems}}}
  \emph{(\bibinfo{series}{{CHI} '21})}. \bibinfo{publisher}{Association for
  Computing Machinery}, \bibinfo{address}{New York, NY, USA},
  \bibinfo{pages}{1--18}.
\newblock
\showISBNx{978-1-4503-8096-6}
\urldef\tempurl%
\url{https://doi.org/10.1145/3411764.3445383}
\showDOI{\tempurl}


\bibitem[Palomares(2008)]%
        {palomares_explaining_2008}
\bibfield{author}{\bibinfo{person}{Nicholas~A. Palomares}.}
  \bibinfo{year}{2008}\natexlab{}.
\newblock \showarticletitle{Explaining {Gender}-{Based} {Language} {Use}:
  {Effects} of {Gender} {Identity} {Salience} on {References} to {Emotion} and
  {Tentative} {Language} in {Intra}- and {Intergroup} {Contexts}}.
\newblock \bibinfo{journal}{\emph{Human Communication Research}}
  \bibinfo{volume}{34}, \bibinfo{number}{2} (\bibinfo{date}{April}
  \bibinfo{year}{2008}), \bibinfo{pages}{263--286}.
\newblock
\showISSN{0360-3989, 1468-2958}
\urldef\tempurl%
\url{https://doi.org/10.1111/j.1468-2958.2008.00321.x}
\showDOI{\tempurl}


\bibitem[Poster(2016)]%
        {poster_5_2016}
\bibfield{author}{\bibinfo{person}{Winifred~R. Poster}.}
  \bibinfo{year}{2016}\natexlab{}.
\newblock \showarticletitle{5. {The} {Virtual} {Receptionist} with a {Human}
  {Touch}: {Opposing} {Pressures} of {Digital} {Automation} and {Outsourcing}
  in {Interactive} {Services}}.
\newblock In \bibinfo{booktitle}{\emph{5. {The} {Virtual} {Receptionist} with a
  {Human} {Touch}: {Opposing} {Pressures} of {Digital} {Automation} and
  {Outsourcing} in {Interactive} {Services}}}. \bibinfo{publisher}{University
  of California Press}, \bibinfo{pages}{87--112}.
\newblock
\showISBNx{978-0-520-96163-0}
\urldef\tempurl%
\url{https://doi.org/10.1525/9780520961630-007}
\showDOI{\tempurl}


\bibitem[Reeves and Nass(1996)]%
        {reeves_media_1996}
\bibfield{author}{\bibinfo{person}{Byron Reeves} {and}
  \bibinfo{person}{Clifford Nass}.} \bibinfo{year}{1996}\natexlab{}.
\newblock \showarticletitle{The {Media} {Equation}: {How} {People} {Treat}
  {Computers}, {Television}, and {New} {Media} {Like} {Real} {People} and
  {Pla}}.
\newblock \bibinfo{journal}{\emph{Bibliovault OAI Repository, the University of
  Chicago Press}} (\bibinfo{date}{Jan.} \bibinfo{year}{1996}).
\newblock


\bibitem[Stumpf et~al\mbox{.}(2020)]%
        {stumpf2020gender}
\bibfield{author}{\bibinfo{person}{Simone Stumpf}, \bibinfo{person}{Anicia
  Peters}, \bibinfo{person}{Shaowen Bardzell}, \bibinfo{person}{Margaret
  Burnett}, \bibinfo{person}{Daniela Busse}, \bibinfo{person}{Jessica
  Cauchard}, \bibinfo{person}{Elizabeth Churchill}, {et~al\mbox{.}}}
  \bibinfo{year}{2020}\natexlab{}.
\newblock \showarticletitle{Gender-inclusive HCI research and design: A
  conceptual review}.
\newblock \bibinfo{journal}{\emph{Foundations and Trends{\textregistered} in
  Human--Computer Interaction}} \bibinfo{volume}{13}, \bibinfo{number}{1}
  (\bibinfo{year}{2020}), \bibinfo{pages}{1--69}.
\newblock


\bibitem[Vorvoreanu et~al\mbox{.}(2019)]%
        {vorvoreanu_gender_2019}
\bibfield{author}{\bibinfo{person}{Mihaela Vorvoreanu}, \bibinfo{person}{Lingyi
  Zhang}, \bibinfo{person}{Yun-Han Huang}, \bibinfo{person}{Claudia
  Hilderbrand}, \bibinfo{person}{Zoe Steine-Hanson}, {and}
  \bibinfo{person}{Margaret Burnett}.} \bibinfo{year}{2019}\natexlab{}.
\newblock \showarticletitle{From {Gender} {Biases} to {Gender}-{Inclusive}
  {Design}: {An} {Empirical} {Investigation}}. In
  \bibinfo{booktitle}{\emph{Proceedings of the 2019 {CHI} {Conference} on
  {Human} {Factors} in {Computing} {Systems}}} \emph{(\bibinfo{series}{{CHI}
  '19})}. \bibinfo{publisher}{Association for Computing Machinery},
  \bibinfo{address}{New York, NY, USA}, \bibinfo{pages}{1--14}.
\newblock
\showISBNx{978-1-4503-5970-2}
\urldef\tempurl%
\url{https://doi.org/10.1145/3290605.3300283}
\showDOI{\tempurl}


\bibitem[West et~al\mbox{.}(2019)]%
        {west_id_2019}
\bibfield{author}{\bibinfo{person}{Mark West}, \bibinfo{person}{Rebecca Kraut},
  {and} \bibinfo{person}{Han Ei~Chew}.} \bibinfo{year}{2019}\natexlab{}.
\newblock \bibinfo{booktitle}{\emph{I'd blush if {I} could: closing gender
  divides in digital skills through education}}.
\newblock \bibinfo{type}{{T}echnical {R}eport}. \bibinfo{institution}{UNESCO}.
\newblock
\urldef\tempurl%
\url{https://docs.edtechhub.org/lib/ZQPSPXQG}
\showURL{%
\tempurl}


\end{thebibliography}

\end{document}